\documentclass[11pt]{article}

\usepackage{tikz}
\tikzstyle{vertex}=[auto=left,circle,fill=black!25,minimum size=20pt,inner sep=0pt]
\usepackage{subfigure}
\usepackage{verbatim}
\usepackage{url}
\usepackage{graphicx, color}
\usepackage{amsmath}
\usepackage{amssymb}
\newcommand{\bs}[1]{\boldsymbol{#1}}
\title{Hierarchical Models for Relational Event Sequences}

\author{Christopher DuBois, Carter T. Butts, Daniel McFarland,
  Padhraic Smyth}

\begin{document}
\maketitle

\begin{abstract}
Interaction within small groups can often be represented as a sequence of events, where each event involves a sender  and a recipient.
Recent methods for modeling network data in continuous time model the rate at which individuals interact conditioned on the previous history of events as well as actor covariates.
We present a hierarchical extension for modeling multiple such sequences, facilitating inferences about event-level dynamics and their variation across sequences. The hierarchical approach allows one to share information across sequences in a principled manner---we illustrate the efficacy of such sharing   through a set of prediction experiments.
After discussing methods for adequacy checking and model selection for this class of models, the method is illustrated with an analysis of high school classroom dynamics.
\end{abstract}

\section{Introduction}

Interactions among small groups of individuals are a focus of study across multiple disciplines.
Investigating the factors that influence group interaction can aid in our understanding of bargaining, leadership, group decision-making, education, and social influence.
For example, considerable attention has been paid to exploring the relationship between group dynamics and task completion \cite{Borgatta1953, bavelas:jacsa:1950}, the formation of norms and attitudes \cite{Festinger1951}, and the creation of roles, status, and hierarchy \cite{Slater1955}.
A key requirement for scientific work in this area is a principled method for analyzing interactions within a specific group as well as a means for comparing phenomena across groups.

Interactions among individuals can often be represented as a sequence of \emph{relational events} -- discrete actions directed from one person to another \cite{Bales1951, Gottman1979, Butts2008}.
In this paper we consider directed, communicative events, though in general an action could be defined as a movement or gesture as well.
Regularities in sequences of  interactions have long been observed empirically across a variety of settings; these include behaviors such as turn-taking according to conversation norms \cite{Gibson2005}, unequal rates of communication acts \cite{Bales1951}, and the influence of contextual factors such as seating arrangement \cite{Hare1963}.
Following the theory that social organization is a product of the actions of individuals \cite{Coleman1990, Simmel1955}, we aim to use sequences of interactions to make inferences about group dynamics.
To achieve this goal we need methods for assessing the explanatory power of theories for individual sequences of events in particular contexts, and for exploring how well such theories generalize across settings.

This paper discusses a family of statistical models for relational event sequences that allow one to study the mechanisms underlying both the sequences of actions and the time until each event.
Our approach builds on recently introduced methods for continuous-time network data \cite{Butts2008} that model the rate at which individuals interact conditioned on the previous history of events, actor covariates, or exogenous events. 

This type of statistical framework allows for parameter estimation, principled model selection, and the incorporation of complex temporal dependencies.  Other types of models also permit inference of temporal dependencies from observational data (e.g. \cite{Parker1988}), but make it difficult to incorporate new mechanisms into the model. 
The framework we employ for modeling relational events has the advantage of being flexible, without sacrificing inferential or computational tractability.

In practice, we often observe multiple sequences.
Collections of sequences may arise for example from multiple trials of a group interaction experiment (in either the same or distinct conditions), from observations from multiple groups or organizational units from a common distribution, etc.  In order to model multiple sequences, we embed the relational event family within a hierarchical Bayesian structure.
This facilitates inferences about event-level dynamics and their variation across sequences, as well as pooling information across a collection of sequences in a principled manner.
An alternative approach for analyzing sequences collected in a variety of situations would be to fit a relational event model to each sequence separately.  Our proposed method instead assumes the parameters are drawn from a common distribution and can thus leverage information from the other sequences when few data are present.
Historically, much (if not most) relational event data has been collected and categorized manually (for example via interaction process analysis \cite{Bales1951}).
Because of the level of effort required, data was often limited to only a few sequences.
With recent technological advances such data is becoming more prevalent, e.g. via mobile devices \cite{Eagle2003}.
The hierarchical approach allows one to compare theories to observed data as well as make predictions about newly obtained sequences obtained in these new settings.

After giving the reader some background on related methods, we review the basic framework for modeling relational event sequences.
We present a hierarchical extension of this framework and discuss methods for fitting such models.
A simulated example is used to illustrate methods for adequacy checking and model selection.
The methodology is then applied to an analysis of high school classroom dynamics from 297 sessions.
We adjust for a variety of actor-level and dyadic covariates, such as role (student or teacher), gender, race, and seating arrangement.
The model specification also includes indicators for typical participation shifts common to human conversation \cite{Gibson2005}, which we explore.
We present estimates from the model and describe their implications.
The paper concludes with a discussion of future directions. 

\section{Background and related work}

Relational structure (and the dynamics thereof) has been treated primarily via a network formalism, in which social actors are conceived of as embedded in a system of temporally extensive relationships.
Traditionally, most social network data have been collected via surveys or by participant observation, with most longitudinal relational data taking the form of \emph{panel data}: a series of ``snapshots'' of an evolving network at various times.
There are several approaches to analyzing this type of data.
Descriptive methods have been used to examine how particular node-level or graph-level statistics change over time, and to describe phenomena such as extent of change in edge structure as a whole (see, e.g., \cite{butts.carley:cmot:2005,butts.cross:joss:2009}).
Statistical approaches, on the other hand, often borrow from methods developed for static networks and parametrize the transition between graphs.
Examples include time-varying latent space methods \cite{Sarkar2005}, latent feature models \cite{Foulds2011}, time-lagged logistic regressions \cite{Almquist2011}, and temporal exponential random graph models (ERGMs) \cite{Wyatt2010,Hanneke2010} using specifications analogous to static ERGMs.
A separate line of work assumes the observed networks are outcomes from a latent continuous time Markov process \cite{Snijders2001, Snijders2005, Snijders2009} and models the switching of  latent edges.

Our research contributes to a growing literature that leverages  continuous-time models from event history analysis  \cite{AalenOddO.2008} in order to model mechanisms at the event level.
Instead of a series of networks at discrete time points, the data is a sequence of edges together with the instantaneous times at which each edge occurred.
In Section \ref{sec:model} we review a framework for modeling for such data, previously applied to observational phenomena such as radio communication during emergency response during the WTC disaster \cite{Butts2008}.
Recent extensions to this approach include modeling  additional real-valued information associated with each event  \cite{Brandes2009}, using approximations to handle multiple-recipient data \cite{Perry2011}, using time-varying coefficients \cite{Vu2011a}, or modeling the temporal dependence using decision trees \cite{Gunawardana2011}.
Other related work assumes that actors have a fixed rate of initiating events and focus modeling efforts on mechanisms affecting the choice of recipient \cite{Stadtfeld2010,Stadtfeld2011}.

\begin{figure*}

\centering
\subfigure[Dynamic network data]
{
\begin{tikzpicture}[scale=.8,transform shape,thick]
  \node[vertex] (n1) at (0,0)  {1};
  \node[vertex] (n2) at (2,2)  {2};
  \node[vertex] (n3) at (4,0) {3};
  \node[vertex] (n4) at (2,-2) {4};
\draw[-stealth] (.4,.4) -- (1.6,1.6);
\draw[-stealth] (2.4,1.6) -- (3.6,.4);
\node at (.7,1.1) {$t_1$};
\node at (3.3,1.1) {$t_2$};
\end{tikzpicture}
}
\subfigure[Intensities for two dyads]
{
\begin{tikzpicture}[scale=1,transform shape,thick]
\draw[-stealth] (0,0) -- (0,3);
\draw[-stealth] (0,0) -- (5,0);

\draw (0,2) -- (1,2);
\draw[fill=black] (1,2) circle (0.75mm);
\draw (1,1.7) circle (0.75mm);
\draw (1.08,1.7) -- (3,1.7);
\draw[fill=black] (3,1.7) circle (0.75mm);
\draw (3,1.7) circle (0.75mm);
\draw (3,2.2) circle (0.75mm);
\draw[-stealth] (3.05,2.2) -- (5,2.2);

\draw[dashed] (0,.3) -- (1,.3);
\draw[fill=black] (1,.3) circle (0.75mm);
\draw (1,1) circle (0.75mm);
\draw[dashed] (1.08,1) -- (3,1);
\draw[fill=black] (3,1) circle (0.75mm);
\draw (3,.75) circle (0.75mm);
\draw[dashed, -stealth] (3.08,.75) -- (5,.75);

\node at (-.5,2) {$\lambda_{1,2}$};
\node at (-.5,1) {$\lambda_{3,4}$};
\node at (1,-.3) {$t_1$};
\node at (3,-.3) {$t_2$};
\end{tikzpicture}
}
\caption[]{Illustration of event data and the assumptions of the model.  Left: An sequence of two events among four nodes: (1,2) occurs at time $t_1$ and (2,3) occurs at time $t_2$.  Right: The intensity functions for  $\lambda_{1,2}(t)$ (solid) and $\lambda_{34}(t)$ (dotted) were chosen to illustrate that each $\lambda_{i,j}(t)$ is a function of the events prior to the last changepoint.}
\label{fig:example}
\end{figure*}
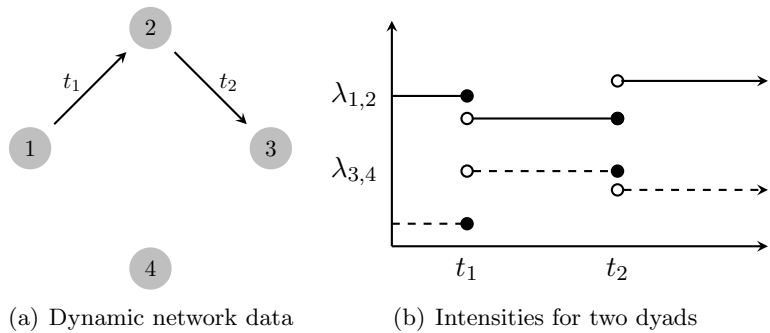

\section{A Framework for Modeling Relational Events}
\label{sec:model}

Suppose we observe a temporally ordered sequence of $M$ dyadic events among $N$ individuals during the time window $[0, \tau) \subset \mathbb{R}^+$,

$$\mathcal{A}_{\tau} = \left\{(t_m,i_m,j_m): (i_m,j_m) \in \mathcal{R}_{t_m}, 0<t_1<\cdots <  t_M < \tau  \right\}$$

where the $m$th event occurs at time $t_m$ and has a sender $i_m$ and a recipient $j_m$.
The set of possible events at time $t$ is referred to as the \emph{risk set}  $\mathcal{R}_t\subseteq\{(i,j): i,j \in \{1,\ldots,N\}\}$; although the risk set can itself be endogenously varying, we here will focus on the case in which $\mathcal{R}_t$ is fixed to the set of all non-reflexive dyadic actions.

Given data of the above form, our primary modeling goal is to understand how the occurrence of a dyadic event $(t,i,j)$ is associated with statistics of the previous history of events as well as dyadic and/or actor-level covariate.
The foundation of this approach borrows from techniques for modeling time-to-event data in survival analysis.
In this context, recall that if $Y$ is a random variable describing the time until an event with probability density function $f$ and cumulative distribution function $F$, then the \emph{survival function} $S(y)=1-F(y) = P(Y>y)$ is the probability of the event not occurring before time $y$, and the \emph{hazard function} $h(y) = f(y)/S(y) = f(y)/(1-F(y))$ is the conditional likelihood that $Y=y$ given $Y \geq y$.
The relational event framework similarly models the time until a given dyadic event occurs, as described below.

An event history $\mathcal{A}$ is modeled as a stochastic process where interactions from node $i$ to node $j$ occur with an intensity $\lambda_{i,j}(t | \mathcal{A}_{t})$ which can depend on previous events.
The likelihood of an observed sequence of $M$ events under noninformative censoring is

\begin{align} \label{eq:genlik}
p(\mathcal{A}_{\tau} | \Theta, X) =& \  \displaystyle\prod_{m=1}^M \lambda_{i_m,j_m} (t_m|\mathcal{A}_{t_m}) \times \notag \\
& \prod_{(i,j) \in \mathcal{R}} S_{i,j}(t_m - t_{m-1}|\mathcal{A}_{t_m}) \times \notag \\
& \prod_{(i,j) \in \mathcal{R}}  S_{i,j}(\tau - t_{M}|\mathcal{A}_{t_M})
\end{align}

\noindent This expression models the fact that $M$ events occurred, that \emph{no other event} occurs between the previous event at time $t_{m-1}$ and the event at time $t_m$, and finally that no events occur between time $t_M$ and $\tau$.
 The non-informative censoring assumption implies that \emph{not} observing event $(i,j)$ during some time interval leaves the waiting time distribution of some other event $(k,l) \in \mathcal{R}$ unchanged.

In order to incorporate dependence on the previous history of events, we follow \cite{Butts2008} and model the hazards $\lambda_{ij}(t | \cdot)$  via a log linear form $$\lambda_{ij}(t|\mathcal{A}_t,\bs{\beta},\mathbf{s},\mathbf{X}) = \exp\{\boldsymbol{\beta}^T \mathbf{s}(t,i,j,\mathcal{A}_t)\}$$ where $\boldsymbol{\beta} \in \mathbb{R}^p$ is a vector of model parameters, and $\mathbf{s}(t,i,j,\mathcal{A}_t)$ is a vector of $P$ statistics pertaining to the dyad $(i,j)$.
The statistics vector $\mathbf{s}(t,i,j,\mathcal{A}_t)$ is a function of  $\mathcal{A}_t$, the event sequence up until time $t$, and covariates about actors and dyads, $\mathbf{X}$.
The future rate of events from $i$ to $j$ can depend on both information that is endogenous (e.g. mixing rates in previous events) and exogenous information (e.g. actor covariates) via the specification of $\mathbf{s}$.
In this approach the intensities $\lambda_{ij}(t | \cdot)$ are piecewise-constant and only can change at event times, as shown in Figure \ref{fig:example}; this automatically satisfies the assumption of non-informative censoring required for Equation~\ref{eq:genlik}.
By construction, the likelihood for a fully observed sequence of events $\mathcal{A}_{\tau}$ then simplifies to

\begin{align}
p(\mathcal{A}_{\tau} | \beta, X) =& \ \displaystyle\prod_{m=1}^M \lambda_{i_m,j_m} (t_m|\mathcal{A}_{t_m}) \times \notag \\
&\prod_{(i,j) \in \mathcal{R}} \exp\left\{ - (t_m - t_{m-1}) \lambda_{i,j}(t_m|\mathcal{A}_{t_m})\right\}  \times \notag \\
&\prod_{(i,j) \in \mathcal{R}} \exp\left\{ - (\tau - t_{M}) \lambda_{i,j}(\tau|\mathcal{A}_{t_M})\right\}
\label{eqn:model}
\end{align}

In subsequent sections we show how to use this framework to model sequences of relational events.
Because all dyadic events are conditionally independent on the last event, one may generate data by repeating the following steps until $t_m$ exceeds the end time $\tau$: 

\begin{enumerate}
\item Given $\mathcal{A}_{t_{m-1}}$ sample the inter-event time: 

$$t_{m} - t_{m-1} \sim \mbox{Exponential}\left(\sum_{i,j} \lambda_{ij}(t_{m-1}|\mathcal{A}_{t_{m-1}})\right)$$

\item Sample the observed event from the categorical distribution 

$$p\left((i_m,j_m) = (i,j)\right) = \frac{\lambda_{i,j}(t_{m-1}|\mathcal{A}_{t_{m-1}})}{\sum_{(i,j) \in \mathcal{R}} \lambda_{i,j}(t_{m-1}|\mathcal{A}_{t_{m-1}})}$$

\end{enumerate}

\noindent In the rest of the paper, we denote a sequence of events generated in the time interval $(0,\tau)$ as $\mathcal{A} |\beta,\mathbf{X} \sim \mbox{REM}(\tau,\beta,\mathbf{s},\mathbf{X})$.

\subsection{Connection to non-homogeneous Poisson processes}

One may also view the above relational event framework as a continuous time Markov jump process whose state space contains all possible dyadic interactions.
The transition matrix (or rate matrix) is modeled via the sufficient statistics $\mathbf{s}$ that are a function of the previous events in the sequence.
The process is completely determined by the times of the transitions and the sequence of states.

One implication of the modeling choices above is that all possible dyadic interactions in $\mathcal{R}$ are Poisson processes which are conditionally independent given the previous event history and external covariates.
This imposes a memoryless property -- the waiting time until the next event cannot depend on the time since the previously observed event -- though more flexibility can be achieved by including additional changepoints.

\subsection{Connection to the Cox proportional hazards model}

If we wish to only model the order of the dyadic events, the likelihood of the event sequence uses the multinomial likelihood: at time $t$ we choose a single dyad $(i,j)$ from among $|\mathcal{R}|$ options with probability proportional to $\lambda_{ij}(t|\mathcal{A}_t)$.
We can rewrite this as follows \cite{Butts2008}:

\begin{align}
p(\mathcal{A}|\bs{\beta}, \mathbf{s}) &= \prod_{m=1}^M \frac{\lambda_{i_m j_m}(t_m|\mathcal{A}_{t_{m}})}{\sum_{(i,j) \in \mathcal{R}} \lambda_{ij}(t_m|\mathcal{A}_{t_{m}})} \nonumber \\
&=\prod_{m=1}^M \frac{\exp\{\bs{\beta}' \mathbf{s}(t_m,i_m,j_m,\mathcal{A}_{t_{m}})\}}{\sum_{(i,j) \in \mathcal{R}} \exp\{\bs{\beta}' \mathbf{s}(t_m,i,j,\mathcal{A}_{t_{m}})\}}
\label{eqn:cox}
\end{align}

This is closely related to the Cox proportional hazards model \cite{Cox1972} with time-varying covariates \cite{Andersen1982}, a regression-based approach from survival analysis for modeling treatment differences using time-to-event data. This approach is semiparametric in nature since only the relative rates are modeled and no parametric assumptions are placed on the baseline hazard.
Here, however, we have 1) $|\mathcal{R}|$ possible events that are ``at risk'' and 2) only one failure time is observed  at each event  while all others are right-censored.
From this view, we have a sequence of local Cox models strung together; the proportional hazards assumption permits inference on the relative rates rather than the baseline hazard.
Note that, as in our construction, this imposes piecewise-constant hazards.

The \emph{competing risks} area of survival analysis \cite{AalenOddO.2008} uses a similar framework to model how the presence of various possible causes are related to the time to an event. Note that since $\mathbf{s}$ includes time-varying covariates and particular events can occur multiple times, relational event models can also be classified as dynamic models for recurrent event data.

\subsection{Connection to multinomial choice models}

Let $R_i(\mathcal{A}_t)$ be the set of possible actions for sender $i$ at time $t$.
Then, conditional on no other event occurring before $i$ acts, the probability that $i$'s next action is $a$ is given by

\begin{align}
p(a|\bs{\beta}, \mathbf{s},\mathcal{A}_t ) &= \frac{\lambda_{i a}(t|\mathcal{A}_{t})}{\sum_{(i,a) \in \mathcal{R}} \lambda_{i a}(t|\mathcal{A}_{t})} \notag \\
&= \frac{\exp\{\bs{\beta}' \mathbf{s}(t,i,a,\mathcal{A}_{t})\}}{\sum_{(i,a) \in \mathcal{R}} \exp\{\bs{\beta}' \mathbf{s}(t,i,a,\mathcal{A}_{t})\}}
\label{eqn:choice}
\end{align}

The above likelihood also exhibits similarities to conditional logit discrete choice models from economics \cite{McFadden1973}.
In this view, the statistics used to model a person's utility can vary over time and are a function of the actions the actor has directed towards others and those directed towards him or her.

\subsection{Process explosion}

When using the proposed approach one must be careful when defining $\lambda_{ij}(t|\mathcal{A}_t) = \exp\{\bs{\beta}' \mathbf{s}(t,i,j,\mathcal{A}_{t})\}$.
Consider a specification of the $p$th effect such that $\beta_p>0$ and $s_p(t,i,j,\mathcal{A}_{t})=N_{ij}(t-)$ where $N_{ij}(t-)$ is the number of occurrences of $(i,j)$ before time $t$.
When $\lambda_{ij}(t|\mathcal{A}_t)$ is large then $(i,j)$ will happen often; in turn $s_p(t,i,j,\mathcal{A}_{t})$ will increase, and this feedback loop might lead to $\lambda_{ij}(t) \rightarrow \infty$ as $t$ approaches some finite time.
For example, \cite{AalenOddO.2008} describes a few particular cases when explosion is expected: 1) if $s_p(t,i,j,\mathcal{A}_{t}) = N_{ij}(t-)^2$  (or any higher order polynomial) and $\beta_p > 0$, 2) if $s_p(t,i,j,\mathcal{A}_{t}) = \exp\{\beta N_{ij}(t-)\}$ with $\beta_p>0$, or 3) if $s_p(t,i,j,\mathcal{A}_{t}) = N_{ij}(t-)/t$ (the mean events per time) and $\beta_p > e^{-1}$.
In general, non-explosion occurs if the expected time until the next event (across all possible states) does not converge to 0; if it does, the overall process will generate events at an increasingly fast pace.

One advantage of the fully parametric  model we pursue is that our estimation occurs with respect to the observed event times, not just the order of events. The estimate of the pacing constant should help model the mean number of events per unit time.
Since our observed data is over a finite time interval, we expect (and will check) that the hazards of the model do not explode within that same time.
In general, process explosion can be avoided by selecting model statistics that remain well-bounded for all $\mathcal{A}_t$ (e.g., those that vary within a finite range).

\section{Hierarchical models for multiple sequences}
\label{sec:hmodel}

In some applications, it is natural to consider the joint analysis of multiple event sequences, e.g., as stemming from experimental replications, observations from multiple groups or organizational units, etc.
Naively, we could proceed by fitting the aforementioned relational event model to multiple sequences individually and pooling the results in some fashion.
However, we may want to share information across sequence models in order to perform population-level inference, understand session-level variation, and stabilize inference for poorly constrained parameters.

To this end, we propose a hierarchical variant of the relational event model.
More specifically, suppose we have $K$ event sequences, denoting the $k$th event history by $\mathcal{A}_k$ with corresponding event covariates $\mathbf{X}_k$, and each sequence is modeled using a distinct parameter vector (i.e., for the $k$th sequence, $\bs{\beta}_k$).
Extending the previous notation, let event history $k$ before time $t$ be denoted as $\mathcal{A}_{k,t}$.
Suppose we model the $k$th sequence via a relational event model with a vector of $P$ effects with corresponding parameters $\bs{\beta}_k \in \mathbb{R}^P$.
We assume $\bs{\beta}_k$ arises from an upper level population distribution such that $\beta_{k,p} \sim \mathcal{N}(\mu_p, \sigma_{p}^2)$, where $\mu_p \in \mathbb{R}$ and $\sigma_p > 0$ are the mean and standard deviation of the upper level distribution for the $p$th effect, respectively.
This added structure enables us to pool information concerning particular effects among sequences.
Thus the final hierarchical model is

\begin{align}
\mathcal{A}_k &\sim \mbox{REM}(\tau_k,\bs{\beta}_{k},\mathbf{s}, \mathbf{X}_k)  & \mu_p &\sim \mbox{Normal}(0,\sigma_{\mu}^2) \\
\beta_{k,p} & \sim \mbox{Normal}(\mu_p,\sigma_p^2) & \sigma_p^2 & \sim \mbox{Inv-Gamma}(\alpha_{\sigma},\beta_{\sigma}) \notag
\end{align}

In the case that a single sequence's parameter overly influences the global value $\mu_p$, a useful variation instead considers a $t$-family such that $(\beta_{k,p} - \mu_p)/\sigma_p \sim t_{\nu}$ where $\mu_p$ is the location, $\sigma_p$ is the scale, and $\nu$ is the degress of freedom.
In this situation we let $\nu_p \sim \mbox{Exp}(r)$ and $r = 1/M$ since an uninformative prior imposes a de facto normality assumption \cite{Geweke1993}.

In the analyses shown here, we place a $N(0,2)$ prior on each $\mu_p$ so that \emph{a priori} 95 percent of the probability mass lies between a multiplicative effect of .01 and 50.
The hyperparameters  $\alpha_{\sigma}$ and $\beta_{\sigma}$ control how close $\beta_{k,p}$ should be to $\mu_p$ when few observations are available.
In practice we use hyperparameters $\alpha_{\sigma}=5$ and $\beta_{\sigma}=1$ so that \emph{a priori} the mean variance is about 0.25; as the amount of data increases, in practice the method is insensitive to these settings.

\subsection{Statistical Inference}
\label{sec:inference}

We are interested in the posterior distribution of our parameters given observed data.
 By applying Bayes rule we can derive the posterior distribution, where $p(\mathcal{A}_k|\tau_k,\bs{\beta}_k,\mathbf{s},\mathbf{X}_k)$ is the likelihood of the lower level model in Equation \ref{eqn:model}:

\begin{align}
p(\beta,\mu,\sigma^2 | \mathcal{A}) \propto& \prod_{k=1}^K p(\mathcal{A}_k |\tau_k,\bs{\beta}_k,\mathbf{s},\mathbf{X}_k)  \notag \\
& \prod_{p=1}^P p(\beta_{k,p} | \mu_p,\sigma_p^2)p(\mu_p)p(\sigma_p^2)
\end{align}

In this section we discuss several methods for obtaining summaries of the posterior distribution.
We begin by optimizing the above posterior and then discuss several techniques for using Markov chain Monte Carlo to better understand the uncertainty in the distribution of the parameters.

\subsubsection{Posterior mode estimation}

One quantity of interest is the posterior mode, or \emph{maximum a posteriori} (MAP) estimates of the model parameters.
When estimating a single event sequence, MAP estimates of $\beta$ allow for regularization in a manner that is not easily obtainable via maximum likelihood estimation.
These estimated may be obtained by iteratively maximizing the log posterior: for each parameter $p$ optimize $p(\mu_p | \{\bs{\beta}\}_{k=1}^K)$ and $p(\sigma_p^2| \{\bs{\beta_k}\}_{k=1}^K, \alpha_{\sigma},\beta_{\sigma})$, then for each sequence $k$ optimize $p(\bs{\beta}_k|\bs{\mu}, \bs{\sigma}^2)$.   Because of the presence of multiple posterior modes with low probability mass this MAP approach is less useful in the hierarchical case than in the case of a single event sequence---appendix \ref{app:posterior-mode} provides further discussion.

\subsubsection{Markov chain Monte Carlo via Metropolis-Hastings}

In the multiple sequence case, we recommend sampling from from full posterior distribution using Markov chain Monte Carlo techniques \cite{Robert2004} rather than seeking point estimates (e.g., via MAP).
Using the Metropolis algorithm we obtain approximate samples from $p(\bs{\beta} | \mathcal{A}, \mathbf{X})$ via a Markov chain whose stationary distribution is the posterior of our parameters.\footnote{We cannot employ Gibbs sampling because the conditional distributions are not in closed form.}
The general procedure involved follows standard practice for Bayesian inferene via MCMC (see, e.g. \cite{Gelman2004}): we draw $N_{burnin} + N_{keep}$ samples, assessing convergence of the Markov chain by a combination of visual inspection of trace plots and formal convergence diagnostics.
Given convergence, we then compute summary statistics (e.g. the posterior mean of our parameters) on the simulated draws in order to summarize relevant properties of the posterior distribution.

One practical difficulty for MCMC simulation with the hierarchical relational event family is that the posterior surface typically has many local modes (as discussed in \ref{app:posterior-mode}).  
While these modes often contain little probability mass, the posterior surface surrounding them can be locally steep; this can result in poor mixing for Metropolis algorithms based on local proposals (e.g., random walk Metropolis), due to the Markov chain ascending a local peak and rejecting moves away from it (thereby becoming ``stuck'' for long periods).  
While careful selection of the proposal distribution (e.g., random walk step size) can obviate this problem, the large numbers of parameters involved in the hierarchical model makes such calibration  difficult.  
To facilitate better mixing, we thus employ an MCMC technique called \emph{parallel tempering} \cite{Geyer1991} (a variation on the swapping algorithm \cite{Madras2003}).

The parallel tempering approach involves simulating $K$ chains in parallel whose acceptance rules vary -- in addition to a base chain constructed in the usual fashion, additional chains are maintained in ``heated'' (i.e., higher-entropy) states that accept proposals more often.
After $t_{swap}$ steps we incorporate an additional Metropolis step that allows for mixing across chains, allowing poor (i.e., low posterior density) states to migrate from ``cooler'' chains to the ``warmer'' chains (where they are readily replaced due to these chains' faster mixing rates).  In effect, the parallel tempering system resembles an adaptive random walk Metropolis-Hastings algorithm, in which local proposals are augmented by longer-range jumps preferentially targeted towards regions of higher posterior density.
We outline the general algorithm below, where the unnormalized density of interest $g(\Theta)$ is the posterior $p(\bs{\beta},\bs{\mu},\bs{\sigma}^2|\mathcal{A},\mathbf{X})$.

\textbf{Algorithm: Parallel Tempering}
In order to sample from an unnormalized density $g(\Theta)$ we create a single Markov chain with an expanded state space by sampling $J$ Markov chains in parallel.

\begin{align*}
\pi(\Theta_1,\ldots,\Theta_J) \propto& \prod_{j=1}^J h_j(\Theta_j) \\
h_j(\Theta_j) \propto & \exp \{-g(\Theta_j)/t_j \}
\end{align*}

\noindent where $t_0 < \cdots < t_J$ and $h_j(\Theta_j)$ is the target distribution for chain $j$.
The parameter $t_j$ tunes the acceptance probability for chain $j$; large values ``flatten'' the posterior distribution making proposals easier to accept (i.e. a ``hot'' chain).
After every $t_{swap}$ draws, we propose a swap between parameters in adjacent chains $j$ and $j+1$ with acceptance probability

\begin{align*}
\min \left\{ 1, \frac{h_j(\Theta_{j+1}^{(t)})}{h_j(\Theta_j^{(t)})}
           \frac{h_{j+1}(\Theta_{j}^{(t)})}{h_{j+1}(\Theta_{j+1}^{(t)}) } \right\}
\end{align*}

Having chains at different temperatures helps one obtain draws that are not stuck in local modes and sample from a wider portion of the posterior.
Such methods have been shown to converge quickly in the presence of isolated, local modes \cite{Madras2003}.

\subsection{Collapsed sampling}
 Since our immediate goal is inference on $\bs{\beta}$, an alternative approach for accelerating convergence  is to integrate out the upper level variance parameter, $\bs{\sigma}^2$, integrating over its uncertainty.
This approach alleviates the issue of peaked modes since these regions contain such small amounts of posterior mass that the sampler is unlikely to find them.

To facilitate integrating out $\sigma^2$ we place an $\mbox{Inv-Gamma}(\alpha_{\sigma},\beta_{\sigma})$ prior on the upper level variance parameter.\footnote{ This prior on the upper level variance parameter may have undesirable qualities when few data exist \cite{Gelman2006}, but here we assume $J$ is not too small (perhaps $>$ 10). In our application $J$ is in the hundreds.}
Because the inverse gamma distribution is a conjugate prior to a Gaussian distribution having a known location parameter, the posterior distribution of $\sigma$ can be written in closed form \cite{Gelman2006}:

\begin{align}
\label{eqn:gibbs.sigma}
\sigma_p^2 &| \{\beta_{k,p} \}_{k=1}^K, \mu_p,  \alpha_{\sigma}, \beta_{\sigma} \sim \notag \\
& \mbox{Inv-Gamma}\left(\alpha_{\sigma} + \frac{K}{2}, \beta_{\sigma} + \frac{1}{2} \sum_{k=1}^K (\beta_{k,p} - \mu_p)^2\right).
\end{align}

The posterior distribution of $\mu$ given $\sigma$ and the lower level parameters can in turn be written as

\begin{align}
\label{eqn:gibbs.mu}
\mu_p | \{\beta_{k,p} \}_{k=1}^K,\sigma_p &\sim \mbox{Normal}\left(\frac{1}{K}\sum_{k=1}^K \beta_{k,p},\frac{ \sigma_p^2}{\sqrt{K}}\right),
\end{align}

which is easily simulated and well-behaved.

In contrast to the above parameters, each $\bs{\beta}_k$ does not have a simple conjugate prior.
In \ref{derivation} we derive the posterior density for each $\beta_{k,p}$ where $\sigma_p^2$ is analytically integrated out.
These densities can be sampled using techniques such as random-walk Metropolis-Hastings, slice sampling \cite{Neal2003}, or Hamiltonian Monte Carlo (HMC) \cite{Neal2010}; we find that using univariate slice sampling for each effect $\beta_{k,p}$ works 
well.

To summarize, sampling from the posterior distribution amounts to the following steps:
\begin{enumerate}
\item For each event history $k$ and effect $p$:
  \begin{enumerate}
    \item Sample $\beta_{k,p} | \mathcal{A}_k, \mu_p, \alpha_{\sigma}, \beta_{\sigma}$  using \ref{derivation}.
  \end{enumerate}

\item For each effect $p$:
  \begin{enumerate}
  \item Sample $\sigma_p^2 |  \{\beta_{k,p} \}_{k=1}^K, \mu_p, \alpha_{\sigma}, \beta_{\sigma}$ using Eqn. \ref{eqn:gibbs.sigma}.
  \item Sample $\mu_p |   \{\beta_{k,p} \}_{k=1}^K, \sigma_p^2$ using Eqn. \ref{eqn:gibbs.mu}.
  \end{enumerate}
\end{enumerate}

\subsubsection{Computational considerations}

Evaluating the likelihood for the hierarchical relational event model can be computationally expensive.
Note that sampling the collection of $\bs{\beta}_k$ parameter vectors can be done in parallel.
Even so, for a given sequence the na\"ive approach has time complexity $O(M \cdot P \cdot N^2)$ when we have $M$ events, $N$ actors, and $P$ covariates.
For model specifications of interest, however, dyads may share similar covariates vectors and many covariate vectors may stay constant for large portions of the event sequence.
We take advantage of this fact to obtain significant computational savings.
During model fitting the computation of the second term becomes $O(P \cdot |U|)$, where $U$ is the set of unique vectors $\mathbf{s}(t,i,j,\mathcal{A}_t)$ across all $i,j,t$.
See \ref{llkcomputation} for details.

\subsection{Model selection}
\label{sec:selection}

The proposed framework has a large amount of flexibility with respect to model specification: for example, one can adjust for sender-specific, receiver-specific, and event-specific covariates.
For model selection we employ the deviance information criterion (DIC), a commonly used criterion that attempts to balance gains in predictive accuracy with a penalty for using a large number of effective parameters.
Given some probability model $p(y|\theta)$ with data $y$ and parameters $\theta$,   the \emph{deviance} is -2 times the log-likelihood, $D(y,\theta) = -2 \log p(y|\theta)$.
One approximation for the \emph{effective number of parameters} is the difference between the posterior mean of the deviance and the deviance of the posterior mean 

$$p_D = \frac{1}{L}\sum_{l=1}^L D(y,\theta^{(l)}) - D(y,\hat{\theta})$$ 

\noindent where $\theta^{(l)}$ is one of $L$ samples from the posterior and $\hat{\theta}$ is a point estimate such as the posterior mean \cite{Spiegelhalter2002}.
The DIC is then defined as 

$$DIC = \frac{1}{L}\sum_{l=1}^L D(y,\theta^{(l)}) + p_D$$

Smaller DIC values are preferred.
The statistic is readily computed using MCMC samples of the posterior and is readily applicable to hierarchical models (whereas other standard approaches, such as AIC and BIC, are not).

\subsection{Prediction}
\label{sec:prediction}

Another approach to evaluate our hierarchical models is via a prediction task.
When developing such models, we are interested in how our models generalize to both future data for a given sequence and to other sequences.
If the hierarchical model is pooling information among the observed sequences, then new sequences from the population should be less surprising under the model.
We use \emph{recall} as a measure of surprise: the rank of the observed event in a sorted list of our predicted outcomes.

Given $K$ event sequences, for $K_{test}$ of the sequences we remove a portion of the sequence to use as test data.
For a given sequence and for each event in the test portion of a sequence, we compute the intensity of each possible outcome in $\mathcal{R}$ using the previous history of events and a vector of parameters sampled from our posterior.
After sorting the predicted intensities of all possible events $\{ \lambda_{i,j}(t_m|\cdot) \}_{(i,j) \in \mathcal{R}}$ in decreasing order, we find the rank of the observed event $(i_m,j_m)$ in the list of predicted intensities, and finally compute recall for a given cutoff $z$ as the proportion of subsequent events that were among the top-$z$ most likely under the model.
Higher values at a given level $z$ indicate that future events were ranked highly by the model, in turn indicating better predictive performance for the model on new data.

\section{Model specification and adequacy checking}

In this section we begin by describing how various hypotheses can be incorporated into the modeling procedure via the specification of the statistics $\mathbf{s}$ and illustrate these techniques through a simulated example.
Once a model is specified it is often of interest to understand the ways in which the current model may be lacking in terms of its ability to explain certain aspects of the observed data.
When using linear regression, for example, one may analyze residuals to check for violations in model assumptions, e.g. non-constant variance or non-linearity with respect to particular effects.
We demonstrate diagnostics for relational event models that leverage both structural information about the aggregate network as well as common time-dependent structures.

\subsection{Incorporating temporal dependencies}
\label{sec:pshifts}

A key benefit of the relational event framework is that one may incorporate effects that capture endogenous social dynamics.
 In particular, we will use effects relating to the ``participation shifts'' of Gibson (2005), which are based on an enumeration of the six possible ways a dyadic interaction can occur (where A, B, X, and Y are distinct actors): AB-BA and AB-BY (called ``turn-receiving''), AB-XA and AB-XB and AB-XY (called ``turn-usurping''), and AB-AY (``turn-continuing'').
For example AB-BA denotes the case where actor A sends to actor B is immediately followed by B sending to A.
Similarly, AB-XB (turn-usurping) denotes the case where (A,B) is followed by (X,B) where X is some other actor.
These participation shifts represent common transitions that can occur during dyadic interaction within small groups.

Per Butts (2008), we represent these types of effects in our model through the use of indicator functions.
If effect $p$ is the statistic representing AB-BA, then we specify $s_{p}(t_m,i,j) = \mathbb{I}((i_m,j_m) = (i,j), (i_{m-1},j_{m-1}) = (j,i))$ to indicate whether the event $(i,j)$ is followed by an event $(j,i)$.
A positive parameter reflects an increased propensity for immediate reciprocation.
The other participation shift effects are defined and interpreted analogously.

In Section \ref{sec:classroom-spec} we will include more complicated temporal structures into the model.
When specifying such statistics it is important to avoid having   dependent terms
 that are near-affine, as this will introduce large variances in the parameter estimates.

\begin{figure*}[th]
  \centering
\subfigure[Observed counts]
{
  \includegraphics[width=.3\textwidth]{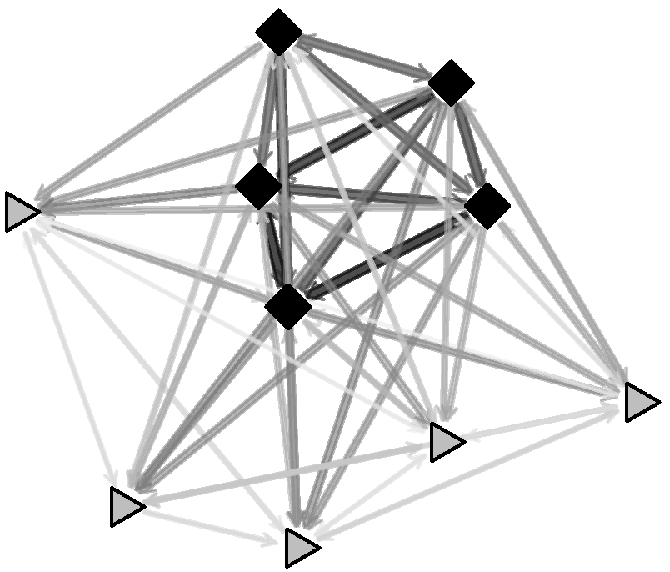}
}
\subfigure[Surprise under Model A]
{
  \includegraphics[width=.3\textwidth]{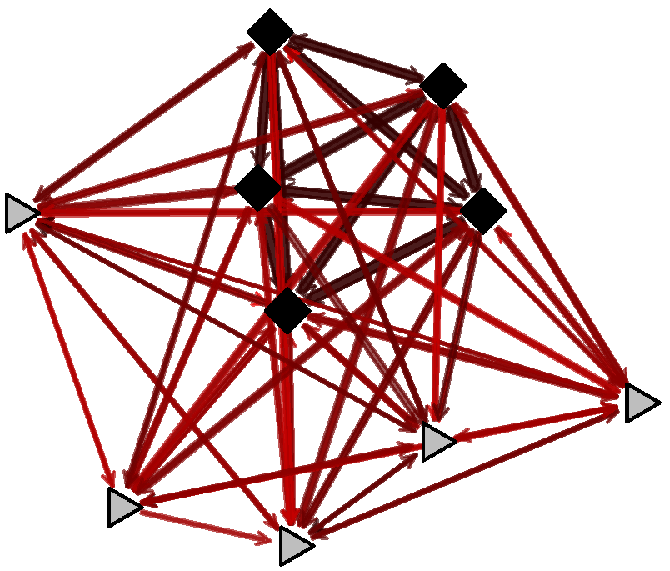}
}
\subfigure[Surprise under Model B]
{
  \includegraphics[width=.3\textwidth]{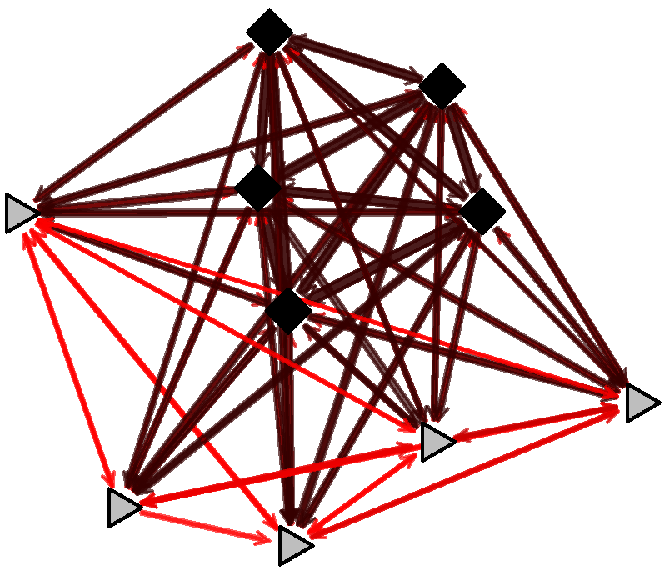}
}
  \caption{Illustration of residual analysis using dyadic marginals for synthetic data.  Edge widths represent number of observations.  Left: Dark colors denote a larger number of events.  Center and right: Red represent a higher amount of surprise under the model (see text).  By adjusting for the observed structure effects for intra-group and inter-group communication, the improved model is only surprised by rare events. }
  \label{fig:dyadicmarginal}
\end{figure*}

\subsection{Synthetic data}
\label{sec:syntheticdata}

To provide an illustration of model specification, consider a synthetic example involving dyadic interactions among 10 individuals. 
Suppose these individuals fall into two classes: triangles and squares.
The dynamics among these actors will be governed by several simultaneous mechanisms, which can be grouped into two categories:
\begin{enumerate}
\item structural: a core-periphery structure whereby squares interact preferentially with other squares, sometimes with triangles, while triangles interact rarely with other triangles
\item time-dependent: an increased propensity for immediate reciprocation, turn-taking, and turn-continuing.
\end{enumerate}

More specifically, we include a multiplicative effect of $e^{1.5}$ for all rates for events among triangles, and $e^{1}$ for events initiated by a triangle and received by a square.\footnote{Note this is done by setting the corresponding parameters $\beta_p$ to 1.5 and 1 respectively; using a loglinear form for the hazard implies the effects are multiplicative.}
Similarly, we include effects that depend on the previous event (say from actor A to actor B): a multiplicative effect of $e^{1.5}$ for the reciprocal event to the previous event (AB-BA), $e^{1}$ for turn-taking events (AB-BY), and $e^{.5}$ for turn-continuing events (AB-AY).
Simulated data from this model can be generated as described in Section \ref{sec:model}.
The network in Figure \ref{fig:dyadicmarginal}a summarizes the synthetic data set, where the edges for commonly occurring dyads have larger width and darker color.
For the examples that follow we simulated 1000 events, of which  we observe 153 AB-AY events, 74 AB-BA events, and 188  AB-BY events.\footnote{Note the probability of this number of AB-BA events happening under a uniform model is Binomial(74;1000,1/100) (very small).}  

\subsection{Dynamic adequacy}

Model adequacy assessment focuses on the basic question of whether (and to what extent) a proposed model effectively captures specified features of an observed data set.  With respect to relational event models, one can consider a wide range of features (both dynamic and temporally aggregated) which may be of scientific relevance in particular situations.  Here, we illustrate some basic examples of heuristics for assessing dynamic adequacy, and for identifying potential avenues for model improvement.

\begin{figure*}[ht]
  \centering
 \includegraphics[width=6in]{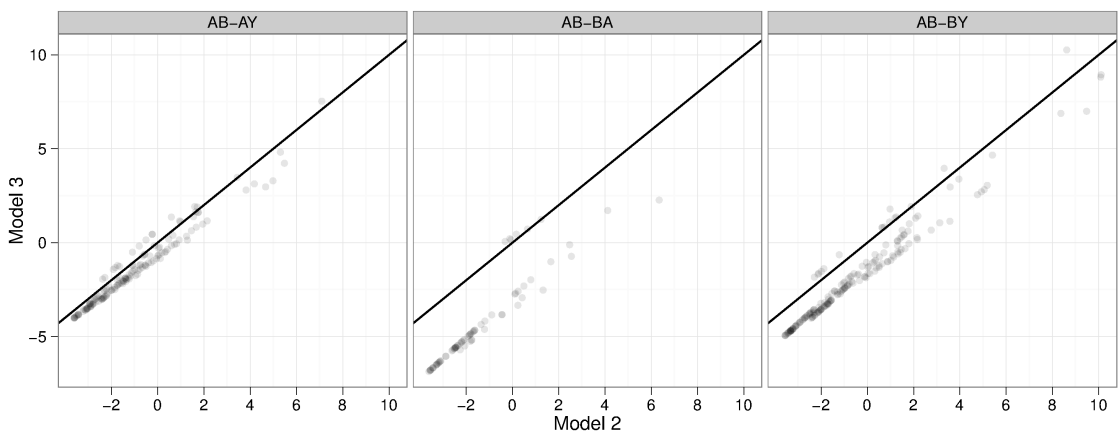}
  \caption{Deviance residuals for events in the synthetic data set.  Lower values are better.  Residuals for each type of event are lower in the model that adjusts for these sorts of temporal effects.}
  \label{fig:continuity}
\end{figure*}

We first consider the \emph{deviance residual} (-2 times the loglikelihood) of each observed event given the previous history of events and a set of model parameters, computed as $$d_m = -2\left[ \log \lambda_{i_m,j_m}(t_m|\cdot) - (t_m - t_{m-1})\sum_{(i,j) \in \mathcal{R}}\log \lambda_{i,j}(t_m|\cdot) \right]$$
Using the synthetic data discussed above, Figure \ref{fig:continuity} compares the deviances for each event under a model with AB-BA effects to a model that also includes AB-BY and AB-AY effects.
Events are categorized by the type of participation shift.
As expected, the set of reciprocated events are better explained once included in the model, indicated by a smaller deviance.  

Alternatively we can consider  the probability of each potential dyadic event being the next to occur in the event sequence, given the previous history of events.  This is equivalent to the likelihood under the order-only model, and is computed as 

$$p((i_m,j_m) | \mathcal{A}_{t_m}) = \frac{\lambda_{i_m,j_m}(t|\cdot)}{\sum_{(i,j) \in \mathcal{R}} \lambda_{ij}(t|\cdot)}.$$

 As noted previously, this can be considered the probability of the observed event under a multinomial probability model.
In Figure \ref{fig:dyadicmarginal2} we compute this probability for models that include participation shifts and those that do not.
Before including them in the model, a number of events with high probability under the true model were given small probability under the fitted model. 

As expected, there are sets of events that are better explained once we include the effects in the model.
In practice, such an analysis can help an analyst better understand \emph{which} events are better explained by the additional effects.
Repeating this analysis for each sequence elucidates how broadly the modeling changes improve fit across the collection of sequences.  

\subsection{Dyadic marginal analysis}

In addition to the intertemporal properties of the event sequence, it is natural to consider the effectiveness of a relational event model in reconstructing the dyadic structure of the time-aggregated interaction network.  It is natural for this purpose to consider the dyadic marginals of the event sequence, wherein we aggregate interactions among directed pairs of individuals across time.  For any given directed pair, $(i,j)$, it is useful to consider the extent to which the observed $(i,j)$ events were considered relatively probable under the proposed model, versus being anomalous.  The presence of pairs for which a large fraction of interactions are ``surprising'' in this sense may suggest the presence of unobserved covariate effects (e.g., homophily), and may provide clues as to the structural context of events for which the model does or does not perform well.

A dyadic measure of ``surprise'' may be computed as follows.
For each event, we rank the predicted probabilities of all possible dyads (ordering randomly for any equal-probability events), and for each dyad compute the proportion of observed events $q_{ij}$ for which the predicted rank exceeds a specified threshold (50, for the synthetic data).
The matrix $q$ may then be viewed as a valued adjacency matrix, such that $q_{ij}$ near 1 reflects that the model is often ``surprised'' by $i,j$ interactions.  By considering $q$ in conjunction with the matrix of raw dyadic event marginals (i.e., the total number of events sent by $i$ to $j$), we can identify portions of the network for which the model is in need of elaboration.

\begin{figure*}
\centering
  \includegraphics[width=5in]{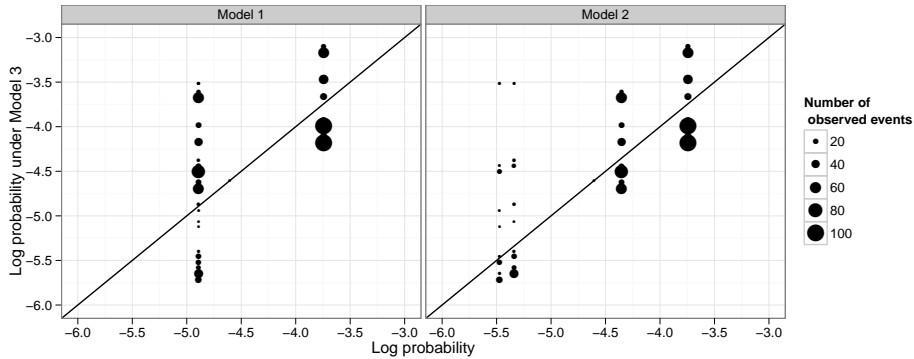}
\caption{Per-event comparison for synthetic data between Models 1, 2, and 3.  Model 3 places adjusting for network structure and temporal structure in Model 3, fewer events are surprising under the model.}
\label{fig:dyadicmarginal2}
\end{figure*}

For our illustrative case, we begin by fitting a model that includes only an effect for events occurring among square nodes, with no other mixing effects. In Figure  \ref{fig:dyadicmarginal}b the edges are colored darker for larger values of $q_{ij}$.
This diagnostic reveals that the fitted model effectively captures communication within the core of square nodes (low surprise), but is surprised by events from the core to the periphery of triangle nodes.
This structural insight suggests incorporating an effect for this set of dyads.
The results of fitting the augmented model and repeating the above procedure are shown in Figure \ref{fig:dyadicmarginal}c.  The square/square and square/triangle interactions are now well-predicted, as shown by the low surprise rate.  Triangle/triangle interactions continue to have high values of $q$, but, as Figure \ref{fig:dyadicmarginal}a shows, these reflect small numbers of rare events that are inherently difficult to predict.  (The total improvement over previous models is shown in Figure \ref{fig:dyadicmarginal2}.)  As this example illustrates, it may be unnecessary (or even impossible) to render all interactions unsurprising under a given model.  Rather, the objective of adequacy assessment is to ensure that important features of the data are well-modeled, and that high levels of surprise are restricted to events that are relatively rare within the data set.

\section{Benefits of the hierarchical approach for multiple event sequences}
\label{sec:benefits}

In this section we illustrate the predictive benefits of the proposed hierarchical approach in the multi-sequence case.
We generate $K=20$ sequences of $M=1000$ events from the model, where $\boldsymbol{\beta}_k \sim N(\boldsymbol{\beta},\sigma^2)$ and $\mathcal{A}_k \sim \mbox{REM}(M,\bs{\beta}_k,\mathbf{s},\mathbf{X})$, where we set $\sigma=1$ and $\bs{\beta}$ and $\mathbf{s}$ are described in Section \ref{sec:syntheticdata}.
 We fit the collection of sequences using 1000 MCMC iterations (see Section \ref{sec:model}).
Through several experiments we show the hierarchical model pools information from other sequences to improve estimation and predictive performance.

\begin{figure*}[th]
  \centering
\subfigure[$\mbox{MSE}(\beta,\hat{\beta}^{ind}) - \mbox{MSE}(\beta,\hat{\beta})$]{
  \includegraphics[scale=.6]{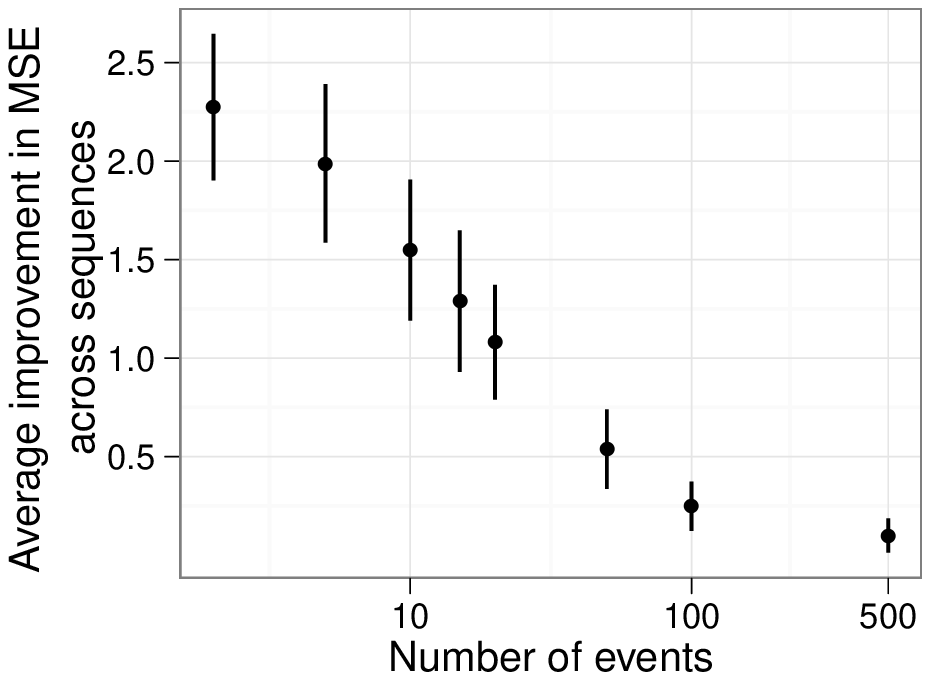}
}
\subfigure[$\mbox{MSE}(\mu,\hat{\mu}) - \mbox{MSE}(\mu,\frac{1}{K} \sum_k \hat{\beta}_k)$]{
  \includegraphics[scale=.6]{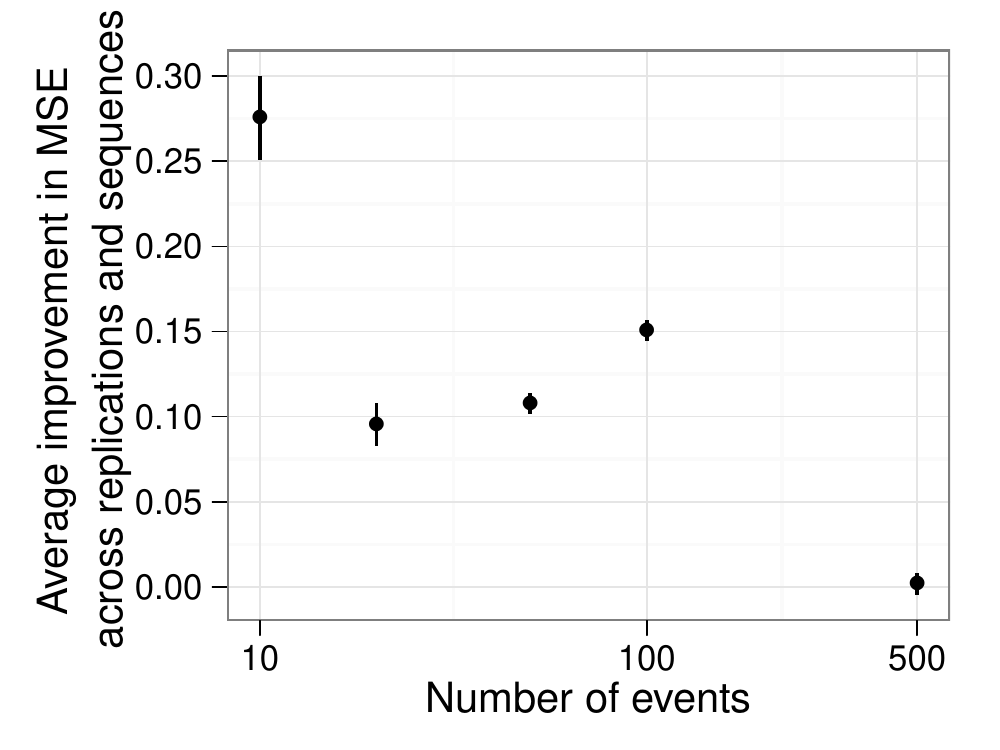}
}
  \caption{Comparing the error in the parameter estimates when using the hierarchical approach compared to fitting separate models $\hat{\beta}^{ind}$. When few events are observed, the hierarchical estimates closer to the truth than the individual fits. See text for details.}
  \label{fig:syn-bias}
\end{figure*}

First we empirically show that, when few events are observed, the hierarchical estimates of the relational event dynamics are more accurate than those obtained by fitting separate models to each sequence, $\hat{\beta}_k$.
To do this, we compare estimates $\hat{\theta}_k$ to the true sequence parameters $\theta_k$ using the mean squared error for that sequence  $\mbox{MSE}(\theta_k,\hat{\theta}_k) = \frac{1}{P}\sum (\hat{\theta}_{kp} - \theta_{kp})^2$.
In Figure \ref{fig:syn-bias} we show the average improvement in MSE across sequences when using the posterior mean from a hierarchical fit, $\hat{\beta}$, versus fitting a model to each sequence separately, $\hat{\beta}^{ind}$.
Bars represent 95 percent confidence intervals of the mean.

We perform a similar experiment with 30 replications of the 20 sequences, now with an emphasis on the benefits in estimating the mean dynamics across the collection of sequences.
We compare the hierarchical approach to simply taking the mean of the lower fits by computing $\mbox{MSE}(\mu,\hat{\mu})$ and $\mbox{MSE}(\mu,\frac{1}{K}\sum_{k=1}^K\hat{\theta}_{k}^{ind})$.
Again, when few events are observed, the hierarchical estimates are less noisy than the individual fits.

Next, we use the synthetic data set to illustrate the technique described in Section \ref{sec:prediction}.
For a given sequence $k$ with $M_{train}$ events, we compute the mean recall at a cutoff $z$ of the next event using various methods.
We compare this to the recall under the true model $\beta$.
  For example, a recall of 65 percent at $z=5$ indicates that this proportion of events are among the 5 most likely under the model given the previous history; higher values are thus better since the observed events are ranked highly by the model.
We include a simple baseline $\hat{p}$ that ranks future dyadic events by the empirical frequency in event history. For the estimates $\hat{\beta}^{ind}$ we make predictions using posterior means from a relational event model fit to each sequence.
We compare these predictions to those made by $\hat{\beta}$, the posterior means from using the hierarchical model, and $\hat{\mu}$, the posterior means of the upper level location parameters.

In Table \ref{tab:synthetic} we compare each of these estimates to estimates of the lower level parameters under the hierarchical model as we vary the amount of training data available $M_{train}$ for the sequence in question.
We consider two recall cutoffs, 5 and 20, to get a sense of how well the model is capturing reciprocity as well as covariate effects.
The results show the hierarchical model pools data effectively: when small amounts of data are available, the hierarchical estimates have similar (or better) predictive performance than the population parameters $\hat{\mu}$; with large amounts of data predicts at least as well as fitting each sequence separately.

\begin{table}[ht]
\begin{center}
\begin{tabular}{lrrrrr}
  \hline
Cutoff $z$ & $M_{train}$ & $\hat{p}$ & $\hat{\mu}$ & $\hat{\theta}^{ind}$ & $\theta^{true}$ \\ 
  \hline
5 & 10 & 29.9 & 0.5 & 12.2 & -5.3 \\ 
   & 50 & 32.8 & 3.3 & 2.2 & -2.2 \\ 
   & 100 & 33.7 & 4.9 & 0.2 & -1.2 \\ 
   & 500 & 34.1 & 6.0 & 0.1 & 0.3 \\ 
  20 & 10 & 11.7 & 6.7 & 8.1 & -5.0 \\ 
   & 50 & 14.9 & 7.9 & 2.3 & -2.2 \\ 
   & 100 & 16.3 & 9.7 & 0.8 & -1.3 \\ 
   & 500 & 18.6 & 10.2 & 0.2 & -0.3 \\ 
   \hline
\end{tabular}
\caption{Average percent improvement in recall of the hierarchical estimates $\hat{\theta}$ across 20 synthetic event sequences.  The results show the hierarchical model pools data effectively: when small amounts of data are available, the hierarchical estimates perform similarly (or better) than the population parameters $\hat{\mu}$; with large amounts of data predicts at least as well as fitting each sequence separately. See text for details.}
\label{tab:synthetic}
\end{center}
\end{table}

\section{Application to dynamics of high school classrooms}

We use the hierarchical model to study student interactions in high school classrooms using data collected via participant observation \cite{McFarland2001}.
The dynamics of interaction in high school classrooms have important consequences for educational research. With increasingly detailed data collection in these settings, we also need to improve our modeling methods so that we can gain insight at the level of individual decisions: in what situations do students stay on task, move off task, or learn better?
A more modest goal is to quantify which factors are associated with the propensity to interact or share information, whether it is the gender diversity of the classroom, the seating chart, or the style of teaching employed, to name just a few.

 The proposed hierarchical method allows us to model interactions among students in a way that 1) adjusts for actor covariates, 2) adjusts for typical participation shifts in conversation, and 3) shares information across hundreds of classroom sessions, enabling inferences at the classroom level and the population level.
After describing the data set, we outline the specification of our models, perform model selection using DIC, and evaluate the predictive ability of the hierarchical approach.

\subsection{Description of the data set}

Each classroom session is represented as a sequence of instantaneous dyadic events, each having a sender, a recipient, and a timestamp.
For the analysis, we select 278 classrooms for which we have data on classroom interactions, the seating chart, and classroom attributes (e.g. subject matter, student grades, and so on).
We do not include sessions for which some of the student attributes (such as gender or race) are missing or unreported.
Information regarding student friendships and shared extracurricular activities is available from survey data administered to these classrooms. 

Each event is coded as occurring during one of three general \emph{contexts}: Lecture, Groupwork, or Silent.
For each of these, one might hypothesize variation in conversational dynamics and norms; we discuss how we account for this in our model below.
 In these high school classrooms lectures are common (lasting for spells ranging between 3 and 40 minutes), containing mostly events that are 1) directed from the teacher to the class as a whole, 2) a question from the student to the teacher, or 3) student-student interactions.
In order to distinguish between events directed at a particular student versus the collective whole, we code all broadcasts as interactions with a representative ``broadcast'' actor.
\footnote{Broadcast events could alternatively be represented a series of dyadic events to each student.  However, this would confound our estimates for dyadic conversational effects.  One idea is to model broadcast events via a entirely separate specification where the broadcast event's dependence on the previous history is modeled via its own parameters.  We leave this to future work.}

\subsection{Model specification}
\label{sec:classroom-spec}

In the specification of each classroom's covariates $\mathbf{X}_k$ we include terms affecting the propensity to send and receive, respectively, using factors such as sex, race, and student/teacher status.
We also include terms indicating when the actors of a dyad are friends, the same race, or the same gender.
These are implemented as indicator variables $x^{ij}_p = \mathbb{I}(x^i_{kp} = x^j_{kp})$ where $x^i_{kp}$ represents the value of covariate $p$ for actor $i$ of sequence $k$.

Some theories posit that the probability of interaction is higher when two individuals have many ``foci'' in common \cite{Feld1981}.
To capture this, for each dyad we include a statistic representing the number of shared extracurricular activities between the two individuals.
A parameter is included for the propensity to interact when two individuals are seating adjacent to each other, since a propinquity effect can be important in small groups \cite{Hare1963}.\footnote{In order to have the ``broadcast'' actor not influence the estimates for gender, student, and race, we use the mean of these statistics across the individuals in the room.}

Two classes of effects are included to specifically model conversational dynamics.
As mentioned in Section \ref{sec:pshifts} we include effects capturing \emph{participation shifts} in the conversation \cite{Gibson2005}.
This represents a collection of first-order Markov effects.
These are likely to be important due to reciprocity effects in general conversation \cite{Gibson2005} and, specifically, in the classroom \cite{McFarland2001}.

For a particular actor, there is reason to expect more recent discussion participants to be more  salient for each actor involved due to mnemonic or contextual factors \cite{Butts2008}.
To model the propensity of an actor to send to his or her recent contacts, we include $j$'s rank in the list of $i$'s most recent in-neighbors and out-neighbors, which we will denote as $\rho_s(i,j,\mathcal{A}(t_m))^{-1}$ where $\rho_r(i,j,\mathcal{A}(t_m))^{-1}$ respectively.
For example, if $j$ is the last person $i$ talked to, $\rho(i,j,\mathcal{A}(t_m))^{-1}=1$; when $i$ talks to a different actor, this value falls to 1/2.\footnote{To ensure that the behavior of $\rho$ is well-defined, actors who do not belong to $i$'s in-neighborhood are considered to have rank $\infty$.}   Both $\rho_s(i,j,\mathcal{A}(t_m))^{-1}$ and $\rho_r(i,j,\mathcal{A}(t_m))^{-1}$ are included in the vector of statistics $\mathbf{s}(t_m,i,j,\mathcal{A}_{t_m})$.

Since the context of an event might modify other effects, for some models we include interaction terms between the event's context and the other statistics.
For example, events associated with groupwork might affect the propensity of teacher-broadcast events, so we consider terms such as $\mathbb{I}(\mbox{Event is Teacher-Broadcast}) \times \mathbb{I}(\mbox{Event is coded ``groupwork''})$.
For these models there are three levels (lecture, groupwork, and silent-time) and we use lecture as the reference level.

\subsection{Model fitting and interpretation}

We fit the hierarchical relational event model to the classroom data described above via MCMC using the techniques in Section \ref{sec:inference}.
A number of model specifications were considered (see \ref{app:specification}).
In Table \ref{tab:classroom-dic} we compare each model's DIC as defined in Section \ref{sec:selection}.
In general we find that including conversational effects such as participation shifts and recency effects provide a large improvement in model fit.
The improvement of B models over the corresponding C models shows that incorporating dyadic covariates helps, e.g. adjacent seating and the number of activities two actors share.  Models that do not include a dyadic effect for teacher-to-class events suffer, as seen by F1 being outperformed by E1 and G1.

\begin{table*}[ht]
\begin{center}
\begin{tabular}{lrrrrrrr}
  \hline
 & A & B & C & D & E & F & G \\ 
  \hline
1 & 416010.97 & 424013.66 & 467672.18 & 438826.80 & 423458.73 & 470573.29 & 436887.00 \\ 
  2 & 491267.67 & 531529.72 & 552627.56 & 555709.71 & 531893.94 & 553670.74 & 560816.36 \\ 
  3 & 544749.70 & 592048.94 & 608186.37 & 619200.81 & 592188.58 & 608604.79 & 618626.41 \\ 
   \hline
\end{tabular}
\caption{DIC under various model specifications for the hierarchical model fit to high school classroom interactions.  Lower is better.  Models with recency statistics tend to be preferred.}
\label{tab:classroom-dic}
\end{center}
\end{table*}

\begin{figure}[htb]
\centering
\includegraphics[scale=.5]{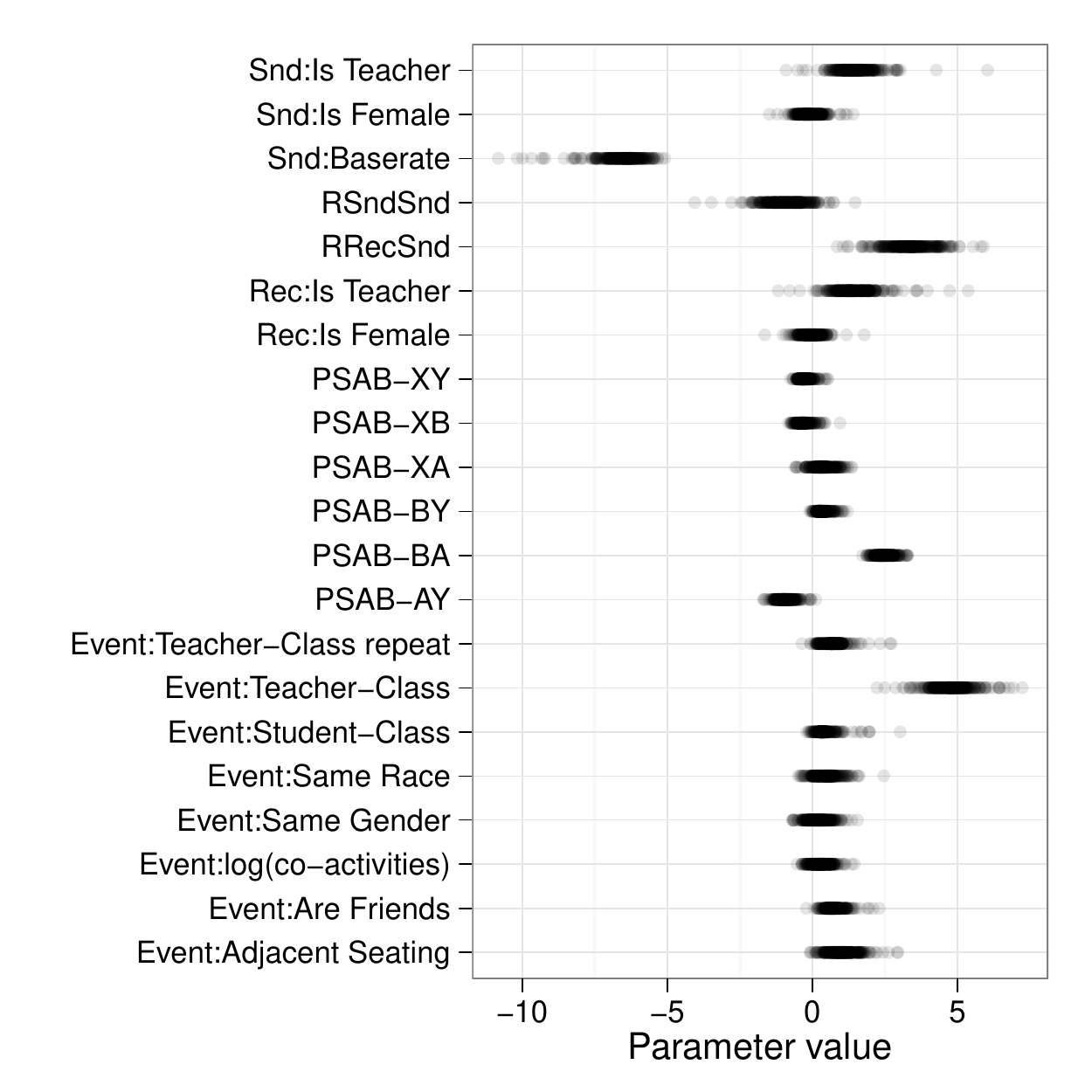}
\label{fig:hierestimates}
\caption{Estimates for each classroom $\hat{\bs{\beta}}_k$.}
\end{figure}

In Figure \ref{fig:hierestimates} we show the individual classroom parameter estimates under Model A1.
The baserate adjusts the overall pacing of the event sequence; a large amount of variability here means that some classroom sessions were slower.
The estimates also reveal a propensity for events to occur from the teacher to the entire classroom (as one might expect), but also positive effects for AB-BA, recency effects, same race, and whether the students are sitting next to each other.

\subsection{Pooling information via the hierarchical model}

\begin{figure}[htb]
\centering
\includegraphics[scale=.5]{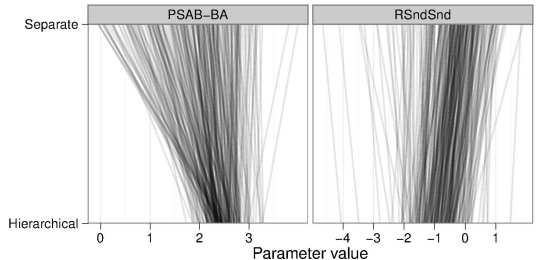}
\caption{Point estimates for a single parameter across all 278 classroom sessions when fitting each session separately (top) and in with the hierarchical model (bottom).  Estimates that were unrealistically high or low (due to problems with estimation) are now informed by estimates from the other sessions and pulled towards the global mean for this model parameter. }
\label{fig:shrinkage}
\end{figure}

In some cases individual sessions have little information concerning certain model effects (e.g. the racial or gender mix within a particular classroom provide few opportunities for cross-group interaction).
Indeed, this is one of the reasons to pursue a hierarchical model: to share information among the classrooms in a principled manner, by ``shrinking'' poorly-informed parameters from individual sessions towards a shared, global value.
We illustrate this for several parameters in Figure \ref{fig:shrinkage} by showing estimates obtained when fitting the relational event model to each classroom separately with respect to the estimates obtained via the hierarchical model.
As expected, uncertain or unconstrained parameters $\beta_{k,p}$ are brought towards the global value $\mu_p$.

\subsection{Prediction experiment}

\begin{figure}
\centering
\includegraphics[scale=.5]{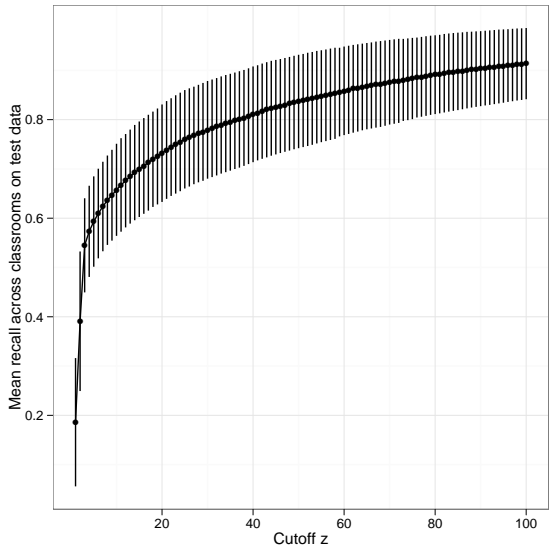}
\label{fig:recall}
\caption{Mean test recall across the high school classrooms.  Estimates obtained from fitting model A-1 with $M_{train}=50$.  Bars show 1 standard deviation.  On average, 60 percent of events are among the fitted model's top 5 most likely events.}
\end{figure}

In this section we evaluate the predictive accuracy of the hierarchical models fit to the collection of high school classroom interactions.  
In Figure \ref{fig:recall} we show the mean recall across classrooms.  
While we observe variability in predictive accuracy across the classrooms, we see that on average 60 percent of subsequent events are among the top 5 most likely under the fitted model.  
This accuracy is partly due to the model capturing the propensity for events involving the teacher and the tendency for reciprocating events.

As in Section \ref{sec:benefits}, the benefits of the hierarchical approach are demonstrated via a comparison to 
  both (a) predictions from fitting separate models and (b) predictions from the population-level parameters.
 For 100 of the sequences we only include the first $M_{train}$ events in the sequence when fitting the model.
The prediction task is to rank the remaining events conditioned on the previous sequence of events (see Section \ref{sec:prediction}).

\begin{table}[ht]
\begin{center}
\begin{tabular}{rrp{2cm}p{2cm}}
  \hline
Cutoff $z$ & $M_{train}$ & $\hat{\mu}$ & $\hat{\theta}^{ind}$ \\ 
  \hline
10 & 50 & 1.8 & 13.4 \\ 
   & 100 & 1.9 & 6.6 \\ 
   & 200 & 1.6 & 4.8 \\ 
  20 & 50 & 1.6 & 12.2 \\ 
   & 100 & 2.6 & 6.1 \\ 
   & 200 & 2.1 & 4.4 \\ 
  40 & 50 & 1.0 & 9.8 \\ 
   & 100 & 1.7 & 4.8 \\ 
   & 200 & 1.5 & 3.7 \\ 
   \hline
\end{tabular}
\caption{Prediction experiment on the classroom data.  Mean improvement in recall across classrooms for varying amounts of training data.}
\label{tab:classroom-recall}
\end{center}
\end{table}

Table \ref{tab:classroom-recall} shows the results of this experiment.
Compared to using separate estimates for sequences of 100 events, the hierarchical model ranks 6.6 percent more observed events among its top 10 most likely and 6.1 percent more observed events among its top 20.

The results show the hierarchical model is able to share information the collection of sequences in a way that improves predictive accuracy on those sequences with missing data.
In the presence of small amounts of data, the accuracy of the model is similar to using the population-level estimates; as more events are observed, the model is able to outperform both the population-level estimates as well as a set
 of models where each is fit to a single sequence.

\section{Discussion}

This paper introduces a hierarchical statistical framework for modeling collections of dyadic event sequences with discussion of parameter estimation, model specification, adequacy checking, and  predictive evaluation.

A central question for the analysis of event sequences often is: how alike or different are the sequences?  In an analysis of high school classroom dynamics, a prediction experiment was used to evaluate the added benefits of modeling the heterogeneity across classrooms.
 Estimates describing the dynamics of the population of sequences had a predictive accuracy that was competitive with other approaches.
Thus, it appears in this case the central tendency appears to be quite dominant and that heterogeneity, though present under the models considered, appears to be less of an issue.
Such inferences are important for understanding these data and future modeling efforts.

The flexibility of the framework allows for a variety of possible extensions.
For example, in some applications one may have actor attributes that change over the course of the event sequence, or warrant more complex time-dependencies than those used here (such as second-order Markov effects \cite{Parker1988}).
The hierarchical structure employed in this paper assumes sequences are exchangeable; in other cases one might want to pool information only within a subset of sequences according to some (known or unknown) attribute of the sequence or those involved.
In such instances one could propose a different prior structure on $\bs{\beta}$ so that information is pooled across individual sequences in a different manner.
In fact one could extend this framework to allow for multilevel modeling whereby the upper level parameters are replaced with a linear predictor incorporating covariates of the event sequence itself.
For example, we may have a vector of covariates $W_k$ that help explain the variation in $\bs{\beta}_{k}$ we might instead assume $\bs{\beta}_{k} \sim \mathcal{N}_P(\bs{\beta}' W_k,I\sigma^2)$.
Exploring other hierarchical models for $\bs{\beta}$ is an interesting future direction.

Fine-grained data from interactions within small groups of individuals is increasingly available, both from online and face-to-face settings.
Statistical inferences from methods such as those proposed here can provide insight into how dynamics of small groups varies across contexts.

\bibliographystyle{elsarticle-num-nourl}
\bibliographystyle{elsarticle-num}
\bibliography{dubois,ctb}

\begin{thebibliography}{10}
\expandafter\ifx\csname url\endcsname\relax
  \def\url#1{\texttt{#1}}\fi
\expandafter\ifx\csname urlprefix\endcsname\relax\def\urlprefix{URL }\fi
\expandafter\ifx\csname href\endcsname\relax
  \def\href#1#2{#2} \def\path#1{#1}\fi

\bibitem{Borgatta1953}
R.~F. {Borgatta, EF and Bales}, {Task and accumulation of experience as factors
  in the interaction of small groups}, Sociometry 16~(3) (1953) 239--252.

\bibitem{bavelas:jacsa:1950}
A.~Bavelas, {Communication Patterns in Task Oriented Groups}, Journal of the
  Acoustical Society of America 22 (1950) 271--282.

\bibitem{Festinger1951}
L.~Festinger, J.~Thibaut, {Interpersonal communication in small groups.},
  Journal of abnormal psychology 46~(1) (1951) 92--9.

\bibitem{Slater1955}
P.~Slater, {Role Differentiation in Small Groups}, American Sociological Review
  20~(3) (1955) 300--310.

\bibitem{Bales1951}
R.~F. Bales, {Interaction Process Analysis}, Addison-Wesleytle, Cambridge, MA,
  1951.

\bibitem{Gottman1979}
J.~M. Gottman, {Detecting Cyclicity in Social Interaction}, Public Health
  86~(2) (1979) 338--348.

\bibitem{Butts2008}
C.~T. Butts, {A Relational Event Framework for Social Action}, Sociological
  Methodology 38~(1) (2008) 155--200.

\bibitem{Gibson2005}
D.~R. Gibson, {Taking Turns and Talking Ties: Networks and Conversational
  Interaction}, American Journal of Sociology 110~(6) (2005) 1561--1597.

\bibitem{Hare1963}
A.~P. Hare, R.~F. Bales, {Seating position and small group interaction},
  Sociometry 26~(4) (1963) 480--486.

\bibitem{Coleman1990}
Coleman, {Foundations of Social Theory}, Harvard University Press, Cambridge,
  MA, 1990.

\bibitem{Simmel1955}
G.~Simmel, {Conflict and the web of group affiliations}, Free Press, Glencoe,
  IL, 1955.

\bibitem{Parker1988}
K.~C.~H. Parker, {Speaking Turns in Small Group Interaction : A
  Context-Sensitive Event Sequence Model}, Journal of Personality 54~(6).

\bibitem{Eagle2003}
N.~Eagle, A.~S. Pentland, {Social Network Computing}, October~(October).

\bibitem{butts.carley:cmot:2005}
C.~T. Butts, K.~M. Carley, {Some Simple Algorithms for Structural Comparison},
  Computational and Mathematical Organization Theory 11 (2005) 291--305.

\bibitem{butts.cross:joss:2009}
C.~T. Butts, B.~R. Cross, {Change and External Events in Computer-Mediated
  Citation Networks: English Language Weblogs and the 2004 U.S. Electoral Cycle
  ∗}.

\bibitem{Sarkar2005}
P.~Sarkar, A.~W. Moore, {Dynamic social network analysis using latent space
  models}, ACM SIGKDD Explorations Newsletter 7~(2) (2005) 31--40.

\bibitem{Foulds2011}
J.~Foulds, C.~DuBois, A.~Asuncion, C.~Butts, P.~Smyth, {A Dynamic Relational
  Infinite Feature Model for Longitudinal Social Networks}, Proceedings of the
  14th International Conference on Artificial Intelligence and Statistics
  (AISTATS) 15 (2011) 287--295.

\bibitem{Almquist2011}
Z.~W. Almquist, C.~T. Butts, {Logistic Network Regression for Scalable Analysis
  of Networks with Joint Edge / Vertex Dynamics}, IMBS Technical Report.

\bibitem{Wyatt2010}
D.~Wyatt, T.~Choudhury, {Discovering long range properties of social networks
  with multi-valued time-inhomogeneous models}, in: AAAI Conference, no.~1,
  2010.

\bibitem{Hanneke2010}
S.~Hanneke, W.~Fu, E.~P. Xing, {Discrete temporal models of social networks},
  Electronic Journal of Statistics 4 (2010) 585--605.

\bibitem{Snijders2001}
T.~A.~B. Snijders, {The statistical evaluation of social network dynamics},
  Sociological Methodology (2001) 361--395.

\bibitem{Snijders2005}
T.~A.~B. Snijders, {Statistical Methods for Network Dynamics}, Work~(1994).

\bibitem{Snijders2009}
T.~A.~B. Snijders, J.~Koskinen, M.~Schweinberger, {Maximum likelihood
  estimation for social network dynamics}, Annals of Applied Statistics (2009)
  1--25.

\bibitem{AalenOddO.2008}
{Aalen, Odd O.}, O.~Borgan, H.~K. Gjessing, {Survival and Event History
  Analysis: A Process Point of View}, Springer, 2008.

\bibitem{Brandes2009}
U.~Brandes, J.~Lerner, T.~A. Snijders, {Networks Evolving Step by Step:
  Statistical Analysis of Dyadic Event Data}, 2009 International Conference on
  Advances in Social Network Analysis and Mining (2009) 200--205.

\bibitem{Perry2011}
P.~O. Perry, P.~J. Wolfe, {Point process modeling for directed interaction
  networks}, New York.

\bibitem{Vu2011a}
D.~Q. Vu, A.~U. Asuncion, D.~R. Hunter, P.~Smyth, {Dynamic Egocentric Models
  for Citation Networks}, Proceedings of the 28th International Conference on
  Machine Learning (2011) 857--864.

\bibitem{Gunawardana2011}
A.~Gunawardana, S.~Meek, F.~Morris, {A Model for Temporal Dependencies in Event
  Streams}, NIPS.

\bibitem{Stadtfeld2010}
C.~Stadtfeld, {Who Communicates with Whom? Measuring Communication Choices on
  Social Media Sites}, 2010 IEEE Second International Conference on Social
  Computing (2010) 564--569.

\bibitem{Stadtfeld2011}
C.~Stadtfeld, A.~Geyer-Schulz, {Analyzing event stream dynamics in two-mode
  networks: An exploratory analysis of private communication in a question and
  answer community}, Social Networks (2011) 1--15.

\bibitem{Cox1972}
D.~Cox, {Regression models and life-tables}, Journal of the Royal Statistical
  Society. Series B (Methodological) 34~(2) (1972) 187--220.

\bibitem{Andersen1982}
P.~Andersen, R.~Gill, {Cox's regression model for counting processes: A large
  sample study}, The Annals of Statistics 10~(4) (1982) 1100--1120.

\bibitem{McFadden1973}
D.~McFadden, {Conditional logit analysis of qualitative choice behavior}.

\bibitem{Geweke1993}
J.~Geweke, {Bayesian treatment of the independent student-t linear model},
  Journal of Applied Econometrics 8~(S1) (1993) S19--S40.

\bibitem{Robert2004}
C.~P. Robert, G.~Casella, {Monte Carlo Statistical Methods}, Springer, 2004.

\bibitem{Gelman2004}
A.~Gelman, J.~B. Carlin, H.~S. Stern, D.~B. Rubin, {Bayesian Data Analysis},
  Chapman \& Hall/CRC, 2004.

\bibitem{Geyer1991}
C.~Geyer, {Markov chain Monte Carlo maximum likelihood}, Computing science and
  statistics: Proceedings of the~(1).

\bibitem{Madras2003}
N.~Madras, Z.~Zheng, {On the swapping algorithm}, Random Structures and
  Algorithms 22~(1) (2003) 66--97.

\bibitem{Gelman2006}
A.~Gelman, {Prior distributions for variance parameters in hierarchical
  models(Comment on Article by Browne and Draper)}, Bayesian Analysis 1~(3)
  (2006) 515--534.

\bibitem{Neal2003}
R.~Neal, {Slice sampling}, Annals of Statistics 31~(3) (2003) 705--767.

\bibitem{Neal2010}
R.~Neal, {MCMC using Hamiltonian dynamics}, in: Handbook of Markov Chain Monte
  Carlo: Methods and Applications, Chapman \& Hall/CRC, 2010, p. 113.

\bibitem{Spiegelhalter2002}
D.~J. Spiegelhalter, N.~G. Best, B.~P. Carlin, A.~van~der Linde, {Bayesian
  measures of model complexity and fit}, Journal of the Royal Statistical
  Society: Series B (Statistical Methodology) 64~(4) (2002) 583--639.

\bibitem{McFarland2001}
D.~A. McFarland, {Student Resistance: How the Formal and Informal Organization
  of Classrooms Facilitate Everyday Forms of Student Defiance}, American
  Journal of Sociology 107~(3) (2001) 612--678.

\bibitem{Feld1981}
S.~Feld, {The focused organization of social ties}, American journal of
  sociology 86~(5) (1981) 1015--1035.

\bibitem{Thibaux2009}
R.~Thibaux, {Efficient implementation of the relational event model and
  extensions}, Tech. rep., University of California, Irvine (2009).

\end{thebibliography}

\appendix
\section{Model specifications}
\label{app:specification}

When specifying the vector of statistics for the classroom data, we group the various dyadic covariates as follows:
\begin{itemize}
\item Group 1: Same Race, Same Gender, Student-Broadcast, Teacher-Broadcast
\item Group 2: Are Friends, Adjacent Seating, log(number of co-activities)
\item Group 3: Teacher-Broadcast, Teacher-broadcast x Previous event was teacher-broadcast
\item Group 4: I(Lecture, Groupwork, Silence)
\end{itemize}

Each model fit in Table \ref{tab:classroom-dic} includes the following sender and receiver effects: Is Teacher, Is Female, Is White.
Each model also includes one or more of the above groups of dyadic covariates, identifying this choice with a letter:
\begin{enumerate}
\item[A.] Group 1 and Group 2 and Group 3
\item[B.] Group 1 
\item[C.] Group 2 
\item[D.] Group 3
\item[E.] Group 1 x Group 4
\item[F.] Group 2 x Group 4
\item[G.] Group 3 x Group 4
\end{enumerate}

The numbers denote the set of conversational effects included in a given model:
\begin{enumerate}
\item Recency effects and participation shifts
\item Participation shifts
\item Recency effects
\end{enumerate}

For example, ``Model B2'' includes sender and receiver effects (Is Teacher, Is Female, Is white); the letter ``B'' indicates that several dyadic effects are included (Are Friends, Adjacent Seating, log(number of co-activities); and finally the number ``2'' indicates that participation shift effects are included.

\section{Derivation of Sampling Equations}
\label{derivation}

Given the hierarchical model $\theta_k \sim \mbox{Normal}(\mu,\sigma^2)$ with prior $\sigma^2 \sim \mbox{Inv-Gamma}(\alpha,\beta)$, we integrate out $\sigma$ as follows:

\begin{align*}
&p(\theta_k | \mu, \alpha, \beta) \\
&\propto \int_{\sigma} p(\theta_k| \mu, \sigma^2) p(\sigma^2|\alpha,\beta) d \sigma^2\\
&=\int_{\sigma} \frac{1}{\sqrt{2 \pi \sigma^2}} \exp\left\{-\frac{(\theta_k - \mu)^2}{2\sigma^2}\right\} \frac{\beta^{\alpha}}{\Gamma(\alpha)} (\sigma^2)^{-(\alpha+1)}\exp\left\{-\frac{\beta}{\sigma^2}\right\}d\sigma \\
&=\frac{1}{\sqrt{2 \pi}} \frac{\beta^{\alpha}}{\Gamma(\alpha)}  \int_{\sigma} (\sigma^2)^{-(\alpha+1+1/2)} \exp\left\{ -\left[ \frac{(\theta_k - \mu)^2}{2\sigma^2}  + \frac{\beta}{\sigma^2} \right] \right\}d\sigma \\
&=\frac{1}{\sqrt{2 \pi}} \frac{\beta^{\alpha}}{\Gamma(\alpha)}  \int_{\sigma} (\sigma^2)^{-(\alpha+1+1/2)} \exp \left\{ -\left[ \frac{(\theta_k - \mu)^2/2 + \beta}{\sigma^2}\right] \right\} d\sigma \\
&=\frac{1}{\sqrt{2 \pi}} \frac{\beta^{\alpha}}{\Gamma(\alpha)} \frac{\Gamma(\alpha + 1/2)}{\left[ (\theta_k - \mu)^2/2 + \beta\right]^{\alpha+1/2}}
\label{eqn:hmc}
\end{align*}



\section{Computing the likelihood}
\label{llkcomputation}

 Let $U$ be the set of unique vectors $\mathbf{s}(t,i,j)$, and define an  indexing such that $U_{(i,j,k)} = \mathbf{s}(t_k,i,j,\mathcal{A}_k)$.  Note for a given tuple $(i,j,k)$ there is only one element of $U$ equal to $\mathbf{s}(t,i,j,\mathcal{A}_k)$, but there are one or more tuples $(i,j,k)$ where $U_r = U_{(i,j,k)}$ for some element $r$.
 We can rearrange the computation of the loglikelihood to take advantage of the set of unique vectors, $U$, and reduce the amount of computation required during MCMC:

\begin{align*}
&\ell(\bs{\beta} | \mathcal{A}_{t_M}) \\
=&\sum_{m=1}^M [\log \lambda_{i_m,j_m}(t_m|\mathcal{A}_{t_m}) - \\
&\log \sum_{(i,j) \in \mathcal{R}}  (t_m - t_{m-1}) \lambda_{i_m,j_m}(t_m|\mathcal{A}_{t_m}) ]\\
 =& \sum_{m=1}^M \sum_{r=1}^{|U|} \mathbb{I}(U_r = U_{(i_m,j_m,k)}) \bs{\beta}' \mathbf{s}(t_m,i_m,j_m,\mathcal{A}_m) - \\
&  \sum_{m=1}^M \sum_{(i,j) \in \mathcal{R}} \sum_{r=1}^{|U|} \mathbb{I}(U_r = U_{(i,j,k)}) (t_m - t_{m-1}) \exp\{ \bs{\beta}' \mathbf{s}(t_m,i,j,\mathcal{A}_m) \} \\
 =& \sum_{r=1}^{|U|}\left[ \sum_{m=1}^M \mathbb{I}(U_r = U_{(i_m,j_m,k)}) \bs{\beta}'U_r \right] - \\
& \sum_{r=1}^{|U|} \left[\sum_{m=1}^M\sum_{(i,j) \in \mathcal{R}} \mathbb{I}(U_r = U_{(i,j,k)}) (t_m - t_{m-1}) \exp\{\bs{\beta}' U_r \} \right]\\
 =& \sum_{r=1}^{|U|}\left[ q_r U_r \bs{\beta} - m_r \exp\{\bs{\beta}'U_r \} \right]\\
\end{align*}

\noindent where $q_r = \sum_{m=1}^MI(U_r = U_{(i_m,j_m,k)})$ and $ m_r = \sum_{m=1}^M \sum_{i=1}^N \sum_{j=1}^N  I(U_r = U_{(i,j,k)}) (t_m - t_{k-1})$.
Note that computing  $q_r$ and $m_r$ is $O(M \cdot N^2)$ but each only needs to be computed once.
Others have considered reordering the computation of these types of expressions \cite{Thibaux2009, Vu2011a}, and the relative benefit of each method depends on the nature of the statistics $\mathbf{s}(t,i,j,\mathcal{A}_t)$.

\section{Posterior mode estimation}
\label{app:posterior-mode}

In the hierarchical setting, we have found that effects with little  information associated with them often have scale parameters whose posterior modes converge to 0 despite having little posterior mass in this region.
Note that the joint posterior has multiple asymptotic modes, since for a particular effect $p$ we can set $\beta_{k,p}$ to be equal across each sequence $k$ and ascend the posterior density surface by moving $\sigma_p$ arbitrarily close to 0.\footnote{This effect is somewhat counterintuitive, since we would expect the posterior distribution of poorly informed parameters to have a large variance due to uncertainty.  Since there is little information in such cases to distinguish parameters from each other across sequences, however, modes exist in which such parameters are approximately equal and the associated scale parameter is approximately zero.}  
Although underappreciated, this is a common behavior of hierarchical models.
For example, \cite{Gelman2004} considers the following simple model that resembles our setup:

\begin{align*}
y_{ij}|\beta_j \sim N(\beta_j,1) & &  \beta_j \sim N(\mu, \tau) & & p(\mu,\tau) \propto 1
\end{align*}

\noindent where $j=1,\ldots,J$ and $i = 1,\ldots,n_j$.
This particular noninformative prior on $\tau$ happens to provide a proper posterior as well as a closed form for the marginal posterior density $p(\tau|y)$.  
It is important to note that problematic modes of the hierarchical joint posterior rarely contain much probability mass, and that they are thus of little inferential importance; their presence, however, makes it difficult for MAP and related methods to yield useful results.

\end{document}